\documentclass{elsart}

\usepackage{amsmath}
\usepackage{amssymb}
\usepackage{graphicx}
\usepackage{latexsym}
\usepackage{aas_macros}
\usepackage{natbib}
\usepackage[squaren]{SIunits}

\newcommand{\scinot}[2]{\ensuremath{#1 \times 10^{#2}}}

\newcommand{\Mstar}[1]{\ensuremath{M_{*}^{#1}}}
\newcommand{\Mplanet}[1]{\ensuremath{M_{\text{p}}^{#1}}}

\addunit{\cm}{\centi\metre}

\addunit{\dyne}{dyn}
\addunit{\erg}{erg}

\addunit{\gauss}{gauss}

\addunit{\persqcm}{\per \cm \squared}
\addunit{\persqcmnp}{\cm \rpsquared}

\addunit{\percubiccm}{\per \cm \cubed}
\addunit{\percubiccmnp}{\cm \rpcubed}

\addunit{\grampersqcm}{\gram \persqcm}
\addunit{\grampersqcmnp}{\gram \usk \persqcmnp}
\addunit{\grampercubiccm}{\gram \percubiccm}
\addunit{\grampercubiccmnp}{\gram \usk \percubiccmnp}

\addunit{\ergpercubiccm}{\erg \percubiccm}
\addunit{\ergpercubiccmnp}{\erg \usk \percubiccmnp}

\addunit{\molpercubiccm}{\mole \percubiccm}
\addunit{\molpercubiccmnp}{\mole \usk \percubiccmnp}

\addunit{\cmpersec}{\cm \per \second}
\addunit{\cmpersecnp}{\cm \usk \reciprocal \second}

\addunit{\cmpersecsq}{\cm \per \second \squared}
\addunit{\cmpersecsqnp}{\cm \usk \second \rpsquared}

\addunit{\gramcmpersec}{\gram \usk \cmpersec}
\addunit{\gramcmpersecnp}{\gram \usk \cmpersecnp}

\addunit{\gramsqcmpersec}{\gram \usk \cm \squared \per \second}
\addunit{\gramsqcmpersecnp}{\gram \usk \cm \squared \second \rpsquared}

\addunit{\yyear}{yr} 
\addunit{\MBU}{MBU} 

\addunit{\jansky}{Jy}

\addunit{\magnitude}{mag}

\addunit{\cmsqpergramnp}{\centi\metre\squared\usk\reciprocal\gram}

\newcommand{\mySun}{\odot}

\addunit{\Msol}{\ensuremath{\mathrm{M}_{\mySun}}}
\addunit{\Rsol}{\ensuremath{\mathrm{R}_{\mySun}}}
\addunit{\Lsol}{\ensuremath{\mathrm{L}_{\mySun}}}
\addunit{\Zsol}{\ensuremath{\mathrm{Z}_{\mySun}}}

\providecommand{\earth}{\oplus}

\addunit{\Mearth}{\ensuremath{\mathrm{M}_{\earth}}}
\addunit{\Rearth}{\ensuremath{\mathrm{R}_{\earth}}}

\addunit{\Mjup}{\ensuremath{\mathrm{M}_{J}}}
\addunit{\Rjup}{\ensuremath{\mathrm{R}_{J}}}

\addunit{\AU}{au}
\addunit{\lightyear}{ly}
\addunit{\parsec}{pc}

\begin{document}

\begin{frontmatter}
\title{Evidence for Growth of Eccentricity and Mass Clearing in a Disc Interior to a Planet}

\author[Rochester]{Richard~G.~Edgar\corauthref{cor}}
\corauth[cor]{Corresponding author}
\ead{rge21@pas.rochester.edu}

\author[Rochester]{Eric~G.~Blackman}
\ead{blackman@pas.rochester.edu}

\author[Rochester]{Alice~C.~Quillen}
\ead{aquillen@pas.rochester.edu}

\author[Rochester]{Peggy~Varni\`{e}re}
\ead{pvarni@pas.rochester.edu}

\author[Rochester]{Adam~Frank}
\ead{afrank@pas.rochester.edu}

\address[Rochester]{Department of Physics and Astronomy, University of Rochester, Rochester, NY 14627}


\begin{abstract}
We present computational results showing eccentricity growth in the inner portions of a protoplanetary disc.
We attribute this to the evolving surface density of the disc.
The planet creates a gap, which adjusts the balance between the 3:1 (eccentricity exciting) and 2:1 (eccentricity damping) resonances.
The eccentricity of the inner disc can rise as high as $0.3$, which is sufficient to cause it to be accreted onto the star.
This offers an alternative mechanism for producing the large holes observed in the discs of CoKu Tau/4, GM Aur and DM Tau.
\end{abstract}



\end{frontmatter}


\section{Introduction}
\label{sec:intro}

Nearly two hundred extra-solar planets have been discovered over the past decade via the radial velocity (RV) method.
Unfortunately, the RV method can only detect planets in tight orbits around older stars, making it unsuitable for detecting planets in the early stages of their formation.
This deprives us of key information about the processes of planetary formation, meaning that many of the most fundamental questions about the formation process remain unanswered.

Recent observations by the \emph{Spitzer} telescope have shown that the accretion discs surrounding some young stars contain large holes \citep{2005ApJ...621..461D,2005ApJ...630L.185C}.
A variety of mechanisms for forming these holes has been proposed, but the most likely involve a planet `damming up' the outer disc while the inner disc dissipates.
The holes have sharp edges, and only a massive planet provides a satisfactory explanation of this.
\citet{2004ApJ...612L.137Q} invoked viscous dissipation of the disc to explain the hole in CoKu Tau/4, while \citet{2006ApJ...640.1110V} suggested that waves launched from Lindblad resonances might be responsible.
In this paper, we present numerical results which suggest another possibility, that of eccentricity excitation.

\section{Numerical Setup}
\label{sec:numerics}

We use the \textsc{Fargo} code of \citet{2000A&AS..141..165M,2000ASPC..219...75M} to perform our calculations.
\textsc{Fargo} is a simple 2D polar mesh code dedicated to disc planet interactions. 
It is based upon a standard, \textsc{Zeus}-like hydrodynamic solver, but owes its name to the \textsc{Fargo} algorithm upon which the azimuthal advection is based.
This algorithm avoids the restrictive timestep typically imposed by the rapidly rotating inner regions of the disc, by permitting each annulus of cells to rotate at its local Keplerian velocity and stitching the results together again at the end of the timestep.
The use of the \textsc{Fargo} algorithm typically lifts the timestep by an order of magnitude, and therefore speeds up the calculation accordingly.
\textsc{Fargo} includes an artifical viscosity like that described by equations~33 and~34 of \citet{1992ApJS...80..753S}, but with $C_2 = 2$ (thereby spreading shocks over two grid zones).
The mesh centre lies at the central star, so indirect terms coming from the planets and the disc are included in the potential calculation.
We make use of an open inner boundary.
If $v_r < 0$ in the innermost active cells, it is copied to the ghost cells.
However, if $v_r > 0$ in the innermost active cells, then the radial velocity in the ghost cells is set to zero.

We use units normalised such that $G=\Mstar{}+\Mplanet{}=1$, while the planet's initial orbital radius is set at $r=1$.
An $\alpha$-type viscosity is assumed, with $\alpha=10^{-4}$ for the physical viscosity.
We assume a constant aspect ratio disc, with $h/r = 0.05$, making the viscous timescale
\begin{equation}
\tau_{\textrm{visc}} = \frac{2 r^2}{3 \nu} = \frac{2 r^2 \Omega}{3 \alpha c_s^2}
\end{equation}
approximately \scinot{4}{5} orbits at $r=1$.
We set the mass ratio $q=\Mplanet{}/\Mstar{}$ to be $10^{-3}$, approximately equal to the Jupiter-Sun value.
The initial disc surface density is $\Sigma \propto r^{-1}$, a reasonable `intermediate' slope which makes the disc-only solution stationary \citep[see][]{1981ARA&A..19..137P}.
The density is set to \scinot{1}{-5} at the planet's radius in our units.
This means that, although the planet is free to migrate, there is insufficient mass present in the disc to cause significant orbital evolution of the planet.
Our introduction of a planet into a smooth disc implicitly assumes that the planet has gone from the gap opening criteria to a Jupiter mass very rapidly.
This is reasonable in both the core accretion \citep{1996Icar..124...62P} and gravitational instability \citep{2003ApJ...599..577B} models.

Our computational domain extends between $r=0.07$ and $r=4$.
It is covered by 128 grid cells in the radial direction (logarithmically spaced) and 384 azimuthally (uniformly spaced).
The effect of the planet's gravitational force is smoothed over 0.6 of the disc thickness.

\subsection{Test without a planet}

\begin{figure}
\includegraphics{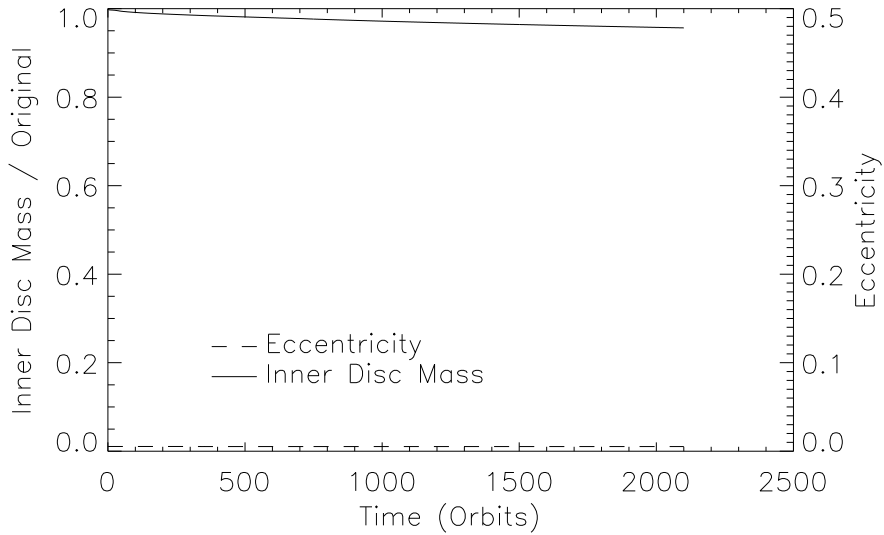}
\caption{Evolution of the mass and mass averaged eccentricity of the inner disc ($0.07 < r < 1$) without a planet. The eccentricity is calculated by assuming that each grid cell is a free particle}
\label{fig:DualEvolveDiscOnly}
\end{figure}

\begin{figure}
\includegraphics{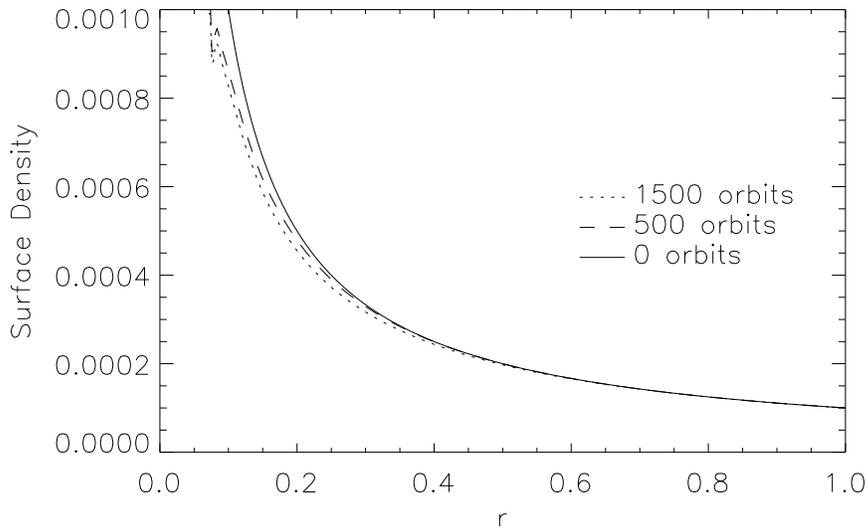}
\caption{Evolution of the disc surface density in the inner disc when no planet is present. The units of surface density are those described in the text}
\label{fig:InnerDiscSurfaceDensityDiscOnly}
\end{figure}

As an initial test, we ran the code without a planet present, but with all other parameters as described above.
Figure~\ref{fig:DualEvolveDiscOnly} shows the evolution of inner disc mass and eccentricity with time, while figure~\ref{fig:InnerDiscSurfaceDensityDiscOnly} gives three snapshots of the surface density.
As expected, the disc is almost stationary.
On examining the viscous timescale as a function of radius, we found it to follow the standard prediction \citep[e.g.][]{1981ARA&A..19..137P}.

\subsection{Eccentricity in the Outer Disc}

\citet{2006A&A...447..369K} recently presented results showing a planet inducing growth of eccentricity in the \emph{outer} disc.
We have run two comparisons to their code, with a $q=10^{-3}$ and $q = \scinot{4}{-3}$ planet.
According to the work of \citeauthor{2006A&A...447..369K}, the Jupter mass planet should not induce significant eccentricity in the outer disc, whereas the \unit{4}{\Mjup} planet should.

\begin{figure}
\includegraphics{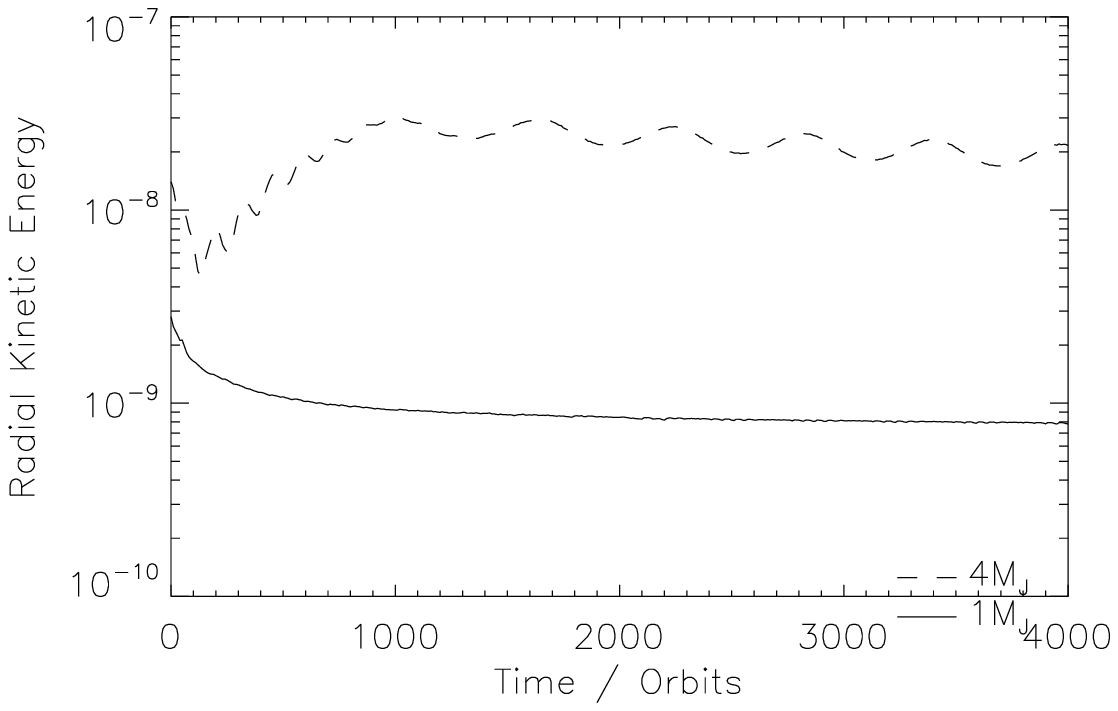}
\caption{Evolution of radial kinetic energy of the disc, as a comparison with \citet{2006A&A...447..369K}. Two different planet masses are plotted}
\label{fig:KDcompareRadialEK}
\end{figure}

\begin{figure}
\includegraphics{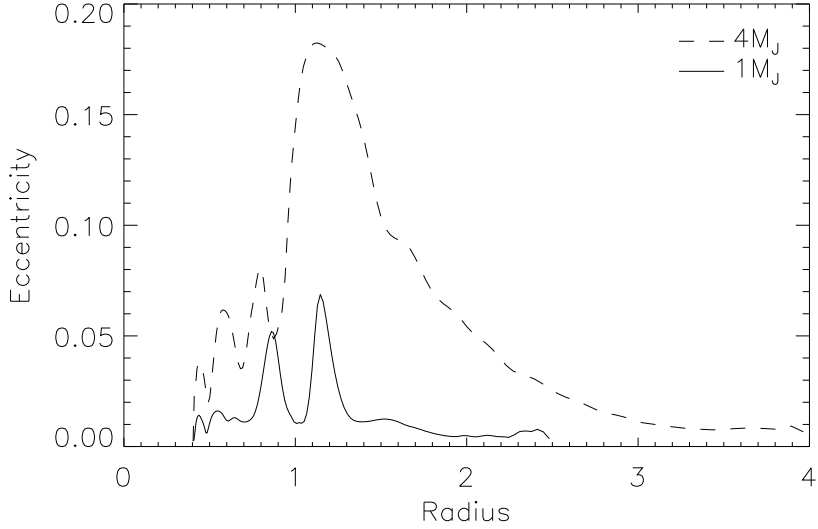}
\caption{Radial eccentricity profiles after 2500 orbits, as a comparison with \citet{2006A&A...447..369K} for two different planet masses}
\label{fig:KDcompareEccProfile}
\end{figure}

Figure~\ref{fig:KDcompareRadialEK} plots the radial kinetic energy of the disc for the two planetary masses.
It should be compared with figure~8a of \citeauthor{2006A&A...447..369K}.
We see that the \unit{4}{\Mjup} planet has excited the eccentricity, whereas the \unit{1}{\Mjup} planet has not.
In figure~\ref{fig:KDcompareEccProfile}, we plot the radial eccentricity profile of the discs, after 2500 orbits.
We see that our results are similar to those of \citeauthor{2006A&A...447..369K}.
In a separate test with a grid extending to $r=2.5$, we found our peak eccentricity to be suppressed, as did \citeauthor{2006A&A...447..369K}.

Although the value of eccentricity reached in our calculation is similar to that observed by \citet{2006A&A...447..369K}, the eccentric mode grew more rapidly in our calculation.
The radial kinetic energy in our test peaks after around 1000 orbits in our test, but after 1500 orbits in the the work of \citet{2006A&A...447..369K}.
We believe that this is due to the effect of the disc mass itself.
\citet{1991ApJ...381..259L} predicted that the eccentricity should grow on a timescale proportional to
\begin{equation}
\lambda \propto q^2 \frac{\Sigma_{\text{res}}}{M_{\text{disc}}}
\label{eq:LubowPredict}
\end{equation}
where the surface density, $\Sigma$, is evaluated at the location of the exciting resonance.
In a linear theory, the density should be proportional to the total disc mass, and hence the disc mass itself should be irrelevant.
However, we are not in the linear regime, and equation~\ref{eq:LubowPredict} takes no account of the damping resonances, or how the eccentric mode propagates through the disc.
\citet{2006A&A...447..369K} imply that their disc mass was actually on the order of the stellar mass, which is unrealistically high.
In further tests, we have found that increasing the disc mass can slow, or even suppress, the growth of eccentricity in the disc.

\section{Results}
\label{sec:results}

Having tested \textsc{Fargo}, we now proceed to our main calculation: that of eccentricity in the inner disc.

\begin{figure}
\includegraphics{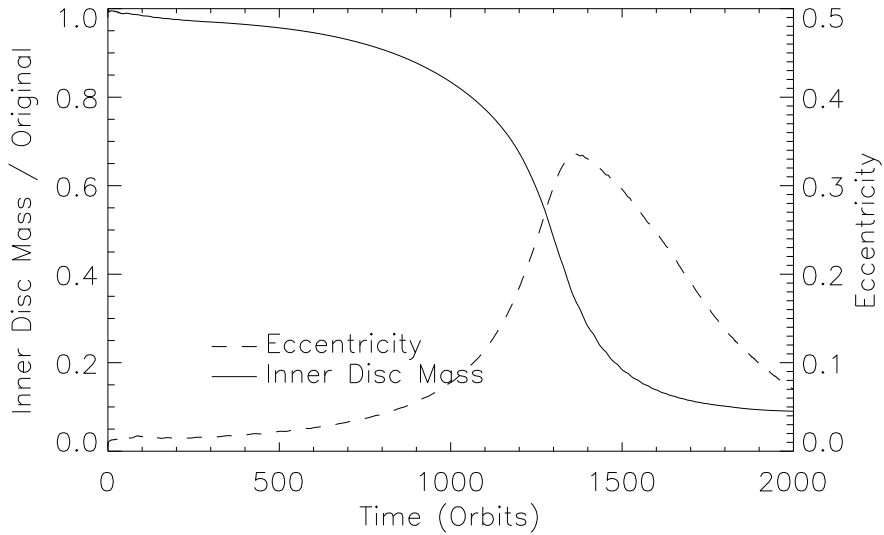}
\caption{Evolution of the mass and mass averaged eccentricity of the inner disc ($0.07 < r < 1$) with a $q=10^{-3}$ planet. The eccentricity is calculated by assuming that each grid cell is a free particle}
\label{fig:DualEvolve01}
\end{figure}

\begin{figure}
\includegraphics{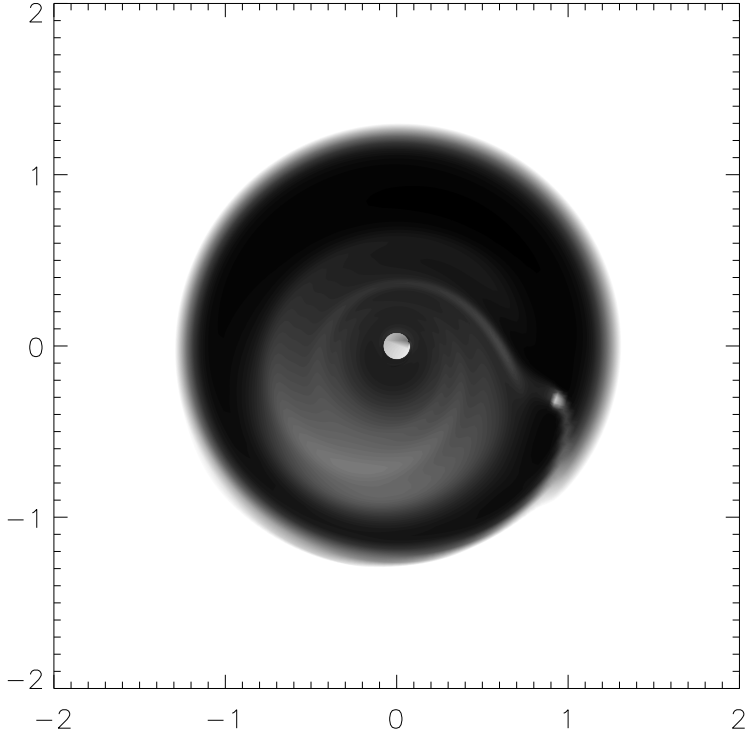}
\caption{Snapshot of the density after 1500 orbits with a  $q=10^{-3}$ planet. The density is plotted relative to the initial value, on a linear scale extending $[0,1]$}
\label{fig:EccDiscImage01}
\end{figure}

In figure~\ref{fig:DualEvolve01} we plot the evolution of the mass and mass averaged eccentricity of the inner disc (between $r=0.07$ and $r=1$).
The eccentricity is quite low for around 500 orbits, when it suddenly starts to climb rapidly.
As its eccentricity grows, the inner disc is depleted (although there is a slight lag behind the eccentricity growth).
We attribute this depletion to material falling onto the central star (i.e. falling off the computational grid).
The eccentricity peaks after around 1500 orbits, and then decays.
The outer disc stays in a low eccentricity state, with $e < 0.02$.
The depletion of the inner disc is far faster than the viscous timescale computed above.
The eccentric inner disc undergoes retrograde precession on a timescale of roughly 80 orbits.
We present a sample snapshot of the density field in figure~\ref{fig:EccDiscImage01}.
The eccentricity of the inner disc is plain in this image.

\begin{figure}
\includegraphics{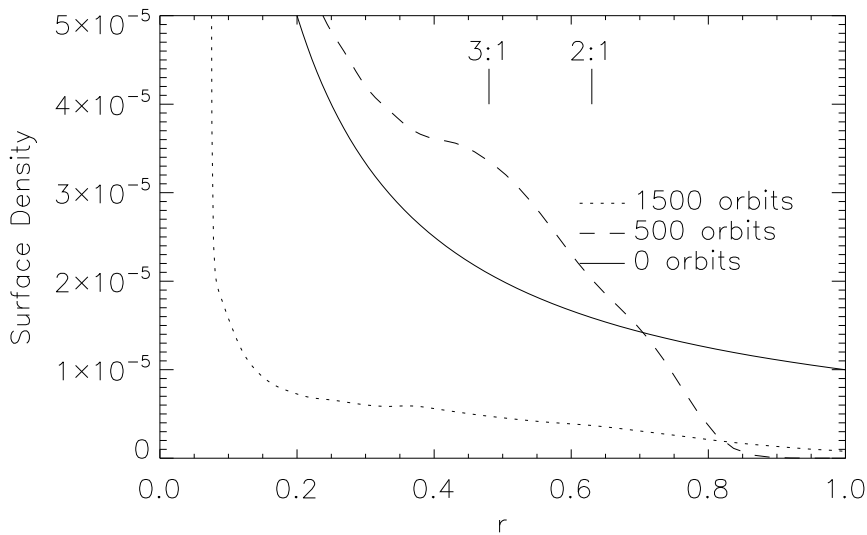}
\caption{Evolution of the disc surface density in the inner disc when a $q=10^{-3}$ planet is present. The units of surface density are those described in the text. The density increase at the inner edge of the disc for the 1500 orbit curve is confined to the two innermost grid cells}
\label{fig:InnerDiscSurfaceDensity01}
\end{figure}

The most likely explanation for this growth lies in the mechanism explored by \citet{1991ApJ...381..259L}.
\citeauthor{1991ApJ...381..259L} describes how a circumstellar disc in a binary star system can become eccentric due to interactions with eccentric Lindblad resonances.
In particular, the eccentric Lindblad resonance located at $r=0.5$ (the 3:1 mean motion resonance) does not concide with an eccentricity damping corotation resonance.
Instead, any eccentricity excited must be damped through the 2:1 corotation resonance (located close to $r=0.6$).
These resonances are finely balanced, and their strengths determined by the local surface density.
In figure~\ref{fig:InnerDiscSurfaceDensity01} we show how the surface density evolves in our calculation.
Between the start of the calculation and 500 orbits, we see that the density at the eccentricity \emph{exciting} 3:1 resonance ($r=0.5$) increases substantially, while the density at the eccentricity \emph{damping} 2:1 resonance ($r=0.6$) increases by a smaller fraction.
This relative increase in excitation enables the 3:1 resonance to `win' and excite the disc eccentricity.
Material is then lost into the hole in the centre of the grid.
As this happens, the density at the 3:1 resonance becomes comparable to that at the 2:1, and damping increases once more.
Plotting eccentricity as a function of radius, we find that the eccentricity peaks close to $r=0.3$, not at the 3:1 resonance itself.
This lends support to the finding of \citet{2006MNRAS.368.1123G} that eccentricity could be suppressed at the location of the exciting resonance.

\begin{figure}
\begin{tabular}{cc}
\includegraphics[scale=0.5]{DualEvolution01.ps} &
\includegraphics[scale=0.5]{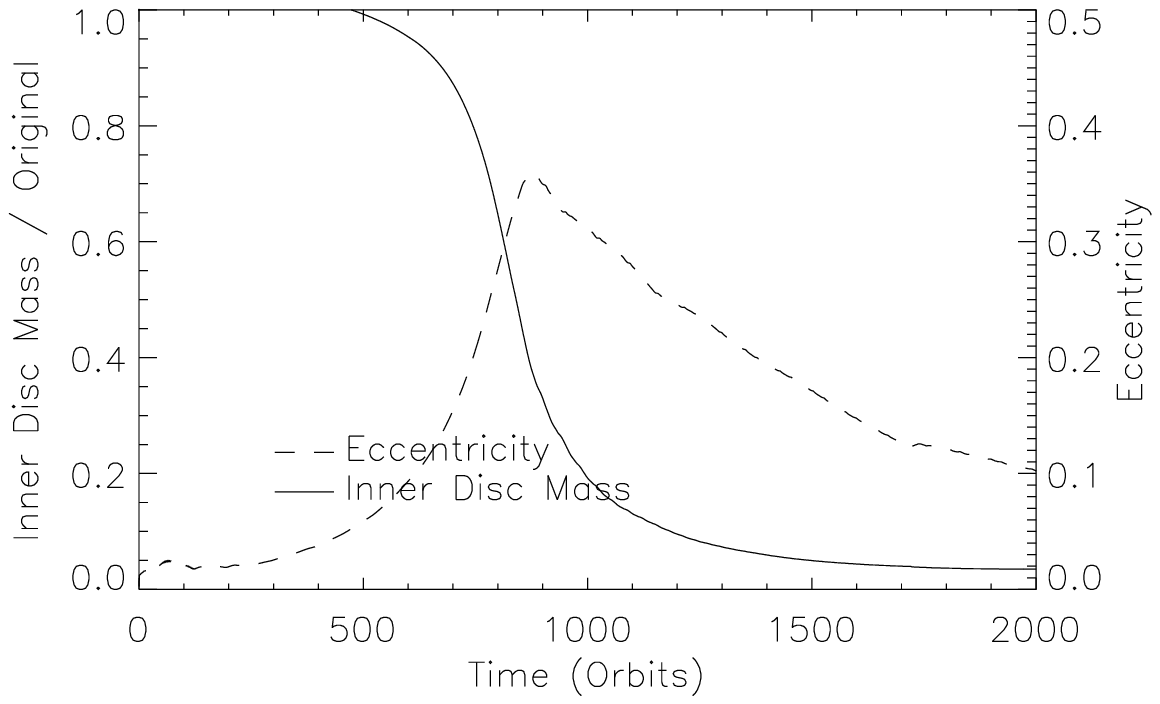} \\
\includegraphics[scale=0.5]{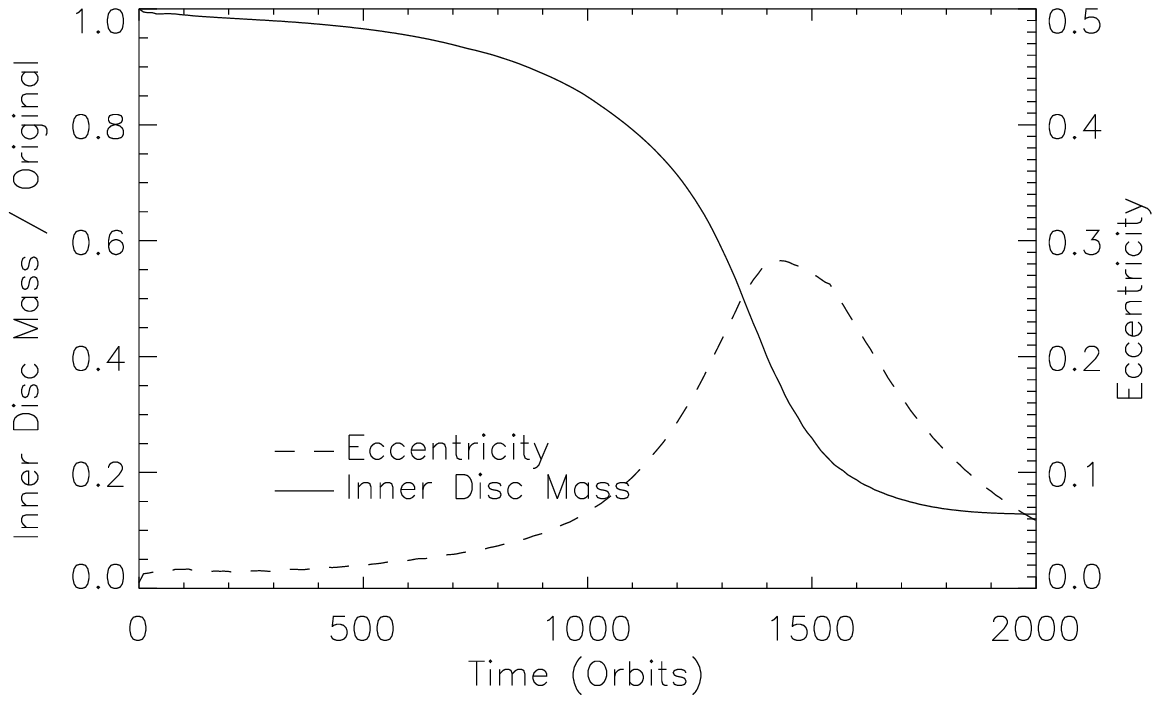} &
\includegraphics[scale=0.5]{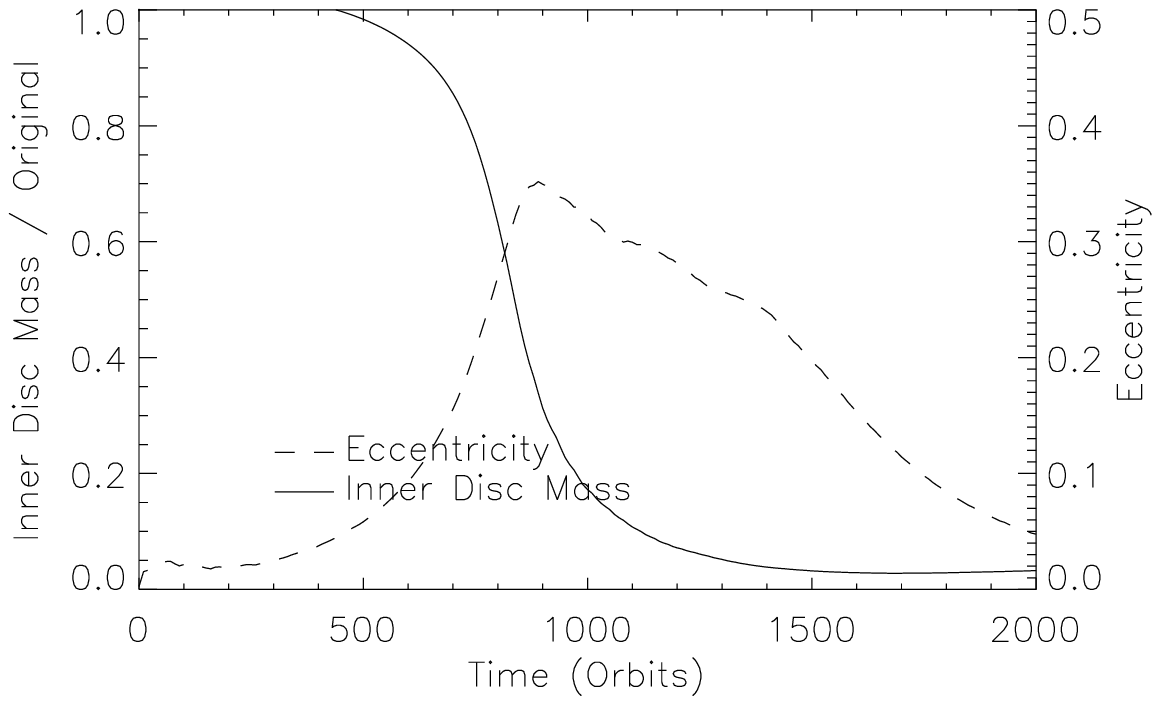}
\end{tabular}
\caption{Evolution of eccentricity and inner disc mass in four runs with a $q=10^{-3}$ planet. Top row: Open inner boundary; Bottom row: non-reflecting inner boundary. Left column: $\Sigma \propto r^{-1}$ initially; Right column: $\Sigma$ initially constant. Hence, the figure in the top left is identical to figure~\ref{fig:DualEvolve01}}
\label{fig:DualEvolveAll}
\end{figure}

In figure~\ref{fig:DualEvolveAll}, we present a set of four runs, each showing growing eccentricity depleting the inner disc.
These vary the initial surface density profile (using $\Sigma \propto r^{-1}$ and constant $\Sigma$) and the inner boundary condition.
Two use the open inner boundary, described above, while two make use of \textsc{Fargo}'s `non-reflecting' boundary.
This attempts to match the guard cells smoothly onto waves propagating on the grid, so that no waves are reflected at the boundary.
We see that in all cases, the eccentricity grew, and depleted the disc.
Changing the boundary had little effect, while changing the intial surface density caused the eccentricity to appear earlier.
A constant surface density is not a steady solution of the viscous disc evolution equation, so viscous evolution will have been occuring simultaneously with the gap opening and eccentricity excitation.

\begin{figure}
\includegraphics{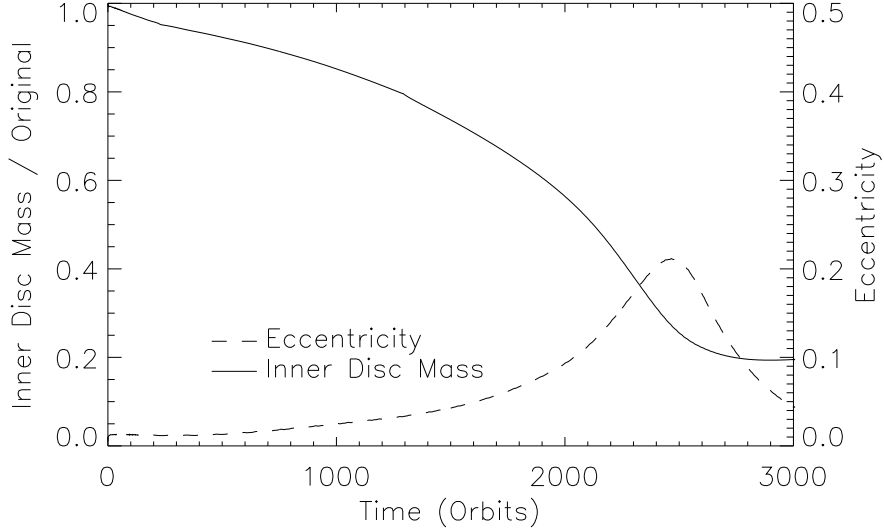}
\caption{Evolution of the mass and mass averaged eccentricity of the inner disc ($0.07 < r < 1$) with a $q=10^{-3}$ planet and $\alpha$ increased to $10^{-3}$. All other parameters were identical to figure~\ref{fig:DualEvolve01}}
\label{fig:DualEvolveHighVisc}
\end{figure}

Figure~\ref{fig:DualEvolveHighVisc} shows the evolution of a disc identical to that in figure~\ref{fig:DualEvolve01} except that it had $\alpha=10^{-3}$ (that is, ten times greater than the assumed viscosity for figure~\ref{fig:DualEvolve01}).
We see that the disc undergoes more viscous evolution (as one would expect), and that the onset of eccentricity growth is delayed.

\citet{2006A&A...447..369K} did not see eccentricity growth in their inner discs because their computational domain was too small.
Their grid only extended to $r=0.25$, which limits the eccentricity which could develop before material would stray from the grid.
Furthermore, their boundary conditions strongly suppressed eccentricity growth in the inner disc.
\citeauthor{2006A&A...447..369K} used `wave killing' boundary conditions \citep[also used in][]{2006MNRAS.370..529D} which push the solution within a `wave killing' region back towards the standard Keplerian disc solution.
Their boundary region extended to $r=0.5$, the location of the 3:1 exciting resonance.
Hence, any eccentricity excited at this resonance would have been subject to strong numerical damping.
Eccentricity excitation of a protoplanet was studied by \citet{2001A&A...366..263P}, but they did not consider the effect on an inner disc.
In a separate test, we found that a grid which only extended to $r=0.2$ did not show such strong eccentricity growth (although computational speed was greatly increased).
Instead, the material in the disc drained steadily, while remaining in near circular orbits.

Explaining the unequal clearing of the resonances is not completely straightforward.
In conventional theory \citep[e.g. the review of][]{2000prpl.conf.1135W}, the disc interacts with the planet at its Lindblad resonances (LR).
Most workers assume local damping, where the angular momentum in the excited wave is immediately transferred to the disc.
In the inner disc, the innermost LR is the $m=2$ which is colocated with the 2:1 resonance.
This would appear to offer an easy explanation: the $m=2$ LR clears material away, weakening eccentricity damping.
There is no similar clearing from the 3:1 exciting resonance, since all other LR lie closer to the planet.
However, in common with many other workers in this area \citep[see][]{2006MNRAS.370..529D}, we see the planet's wake propagating far into the inner disc, indicating that damping is \emph{not} completely local.
Some of the wave may be damped immediately upon launch, but a noticeable fraction remains to propagate through the disc.
It is possible that a sufficient portion of the wave is damped at the resonance to clear the 2:1 damping resonance, but we cannot state this with certainty from our present calculations.
We note that the imposition of a locally isothermal equation of state prevents waves from steepening into shocks as they propagate \citep[cf][]{2002ApJ...569..997R}.

\section{Discussion}
\label{sec:discuss}

Disc eccentricity in an inner disc has been invoked to explain the `superhump' phenomenon in binary stars \citep[e.g.][]{2001MNRAS.324L...1T}.
Indeed, this was the original motivation for the work of \citet{1991ApJ...381..259L,1991ApJ...381..268L}
Scaling typical results according to equation~\ref{eq:LubowPredict}, we find that the results presented above have a surprisingly rapid growth rate for the eccentricity.
However, there are reasons why a straightforward scaling may not be appropriate.
Firstly, there is the issue (mentioned above) of the disc mass.
Our results are certainly affected by the disc mass, while equation~\ref{eq:LubowPredict} (based on a linear calculation) predicts that this should drop out of the growth rate.
However, a Jupiter mass planet opens a gap in the disc on a timescale of 100 orbits or less, thereby putting the system into the non-linear regime.
Secondly, most work on binary stars uses the smoothed particle hydrodynamics (SPH) scheme to model the behaviour of the disc.
SPH is known to possess a high numerical dissipation, and this could well affect the growth rate of the eccentric mode.
The differences between grid-based and SPH-based calculations of planets in discs is vividly demonstrated by figure~10 of \citet{2006MNRAS.370..529D}.
Being performed over a decade later, these SPH codes used nearly two orders of magnitude more particles than \citeauthor{1991ApJ...381..268L} (albeit over a larger disc), and yet still obviously have far higher numerical dissipation than the grid based codes.
In figure~\ref{fig:DualEvolveHighVisc}, we have shown how increasing the viscosity delays the onset of eccentricity growth.
Finally, binary stars typically have much larger mass ratios.
Once $q>0.2$, the 3:1 exciting resonance no longer lies within the stable disc, and eccentricity growth is suppressed.
Although work on superhumps invariably uses a smaller mass ratio (to place the resonance within the disc), other resonances from the secondary will be far stronger in such calculations, and it is not clear what their effect will be.

The growth of eccentricity demonstrated in this paper has important implications for the possibility of hole growth in protoplanetary discs.
Accretion discs containing holes have been observed in three systems: CoKu Tau/4 \citep{2005ApJ...621..461D}, GM Aur and DM Tau \citep{2005ApJ...630L.185C}.
It has already been suggested \citep{2004ApJ...612L.137Q} that CoKu Tau/4 contains a planet.
Other models have been suggested, for example the photoevaporation model of \citet{2006MNRAS.369..216A,2006MNRAS.369..229A}, and there are presently insufficient data to rule all of these out.
However, in CoKu Tau/4 at least, the edge of the accretion disc is observed to be very sharp (from SED fitting), which lends itself to interpretation as the edge of a gap induced by a planet.
The 1000-2000 orbit timescale found in the calculation of section~\ref{sec:numerics} fits easily within the estimated ages of CoKu Tau/4, GM Aur and DM Tau.

Consider the case of CoKu Tau/4:
CoKu Tau/4 is a comparatively young ($\sim \unit{1}{\mega\yyear}$), low mass ($\Mstar{} \sim \unit{0.5}{\Msol}$) system with a large inner hole in its accretion disc.
This hole stretches out to a radius of $\sim \unit{10}{\AU}$, and is depleted by a factor of about $10^5$ in dust.
There is no accretion signature, suggesting that little gas is present in the hole.
Numerical experiments find that outer disc lies a couple of Hill radii from the planet \citep[e.g.][]{2006MNRAS.370..529D}, so a planet with a mass ratio $10^{-3}$ would be in orbit at around \unit{8}{\AU}, with an orbital timescale of approximately \unit{30}{\yyear}.
Two thousand orbits would pass in less than \unit{\ensuremath{10^{5}}}{\yyear}, comfortably less than the age of the system.
However, the formation time of the planet, plus the extra time necessary to produce a $10^{-5}$ depletion in surface density (around ten exponential depletion times) should be borne in mind.

The constraints are less severe for GM Aur.
This is a \unit{1.2}{\Msol} star, which is still accreting some gas.
There is an optically thick outer disc, truncated at \unit{24}{\AU}.
However, there is extra emission from the inner regions, consistent with an optically thin inner dust disc, extending out to around \unit{5}{\AU}.
\citet{2000ApJ...545.1034S} estimate the age as \unit{3}{\mega\yyear}.
If the disc edge is being maintained by a $10^{-3}$ mass ratio planet, we would expect it to be located at about \unit{22}{\AU}, with an orbital period of \unit{94}{\yyear}.
The presence of the optically thin inner disc would be consistent with our timescale.
If we assume that the planet will have formed in around \unit{1}{\mega\yyear}, ten such timescales remain with the age of the system.
The eccentricity growth rate is expected to scale with the square of the mass ratio \citep[equation 66 of][]{1991ApJ...381..259L}.
Putting this into the timescale constraints, we conclude planet with only one third the mass ratio shown above (i.e. \scinot{5}{-4}) would probably be sufficient to produce this system.

It is possible that we are observing an evolutionary sequence, with GM Aur being the `youngest' system and CoKu Tau/4 being the `oldest'.
We place `oldest' and `youngest' in quotes, since they refer to number of orbits completed by the planet, rather than the absolute age.
Further observations of these systems will be useful in constraining their properties.
Recently, \citet{2006ApJ...637L.133E} managed to resolve the inner edge of the accretion disc hole in the TW Hydrae system.
Hopefully, similar techniques could be used on these three young systems.

\section{Conclusion}

In this paper, we have reported a new mechanism for producing a hole in an accretion disc: eccentricity excitation by a protoplanet.
Scaling our results, we have shown that the holes in the discs of CoKu Tau/4, DM Aur and GM Tau can all be explained by the presence of a $\sim \unit{1}{\Mjup}$ planet in these systems.
Future work should analyse the clearing mechanism in more detail, exploring the effects of varying viscosity and planet masses.
Performing a parallel study with a separate code will be extremely useful.
Although we have no reason to believe that \textsc{Fargo} is not performing well, the work of \citet{2006MNRAS.370..529D} underlines how different codes can give subtly different answers to the `same' problem.

In general, we expect a hole to form so long as a gap can form \citep[$q > 40 \mathcal{R}^{-1}$, see e.g.][]{1999ApJ...514..344B} and the disc is insufficiently massive to cause significant migration (roughly, $M_{\textrm{disc}} < \Mplanet{}$).
We also require that the viscous timescale is longer than the timescale for eccentricity growth (this requires $\alpha \lesssim 10^{-2}$), or viscous clearing will dominate.
If the gap opening criterion is fulfilled, the planet will then prevent the outer disc replenishing the inner, regardless of the mechanism which clears the inner disc.
In this paper, we have concentrated on the growth of eccentricity, but viscous clearing and Lindblad waves are alternatives.
We shall explore these in future work.
The recent improvements in observations provided by \emph{Spitzer} open up exciting possibilities in the field of planet formation.
There is now a real possibility that we will be able to observe planetary systems in the process of formation.
This will provide strong constraints on theoretical and numerical predictions such as ours.


\section*{Acknowledgements}

We acknowledge support from NSF grants AST-0406799, AST-0098442, AST-0406823, and NASA grants ATP04-0000-0016 and NNG04GM12G (issued through the Origins of Solar Systems Program).
We are grateful to Gordon Ogilvie for information about eccentric accretion discs.
We thank Frederic Masset for use of the \textsc{Fargo} code.
Some of the computations presented here used the resources of HPC2N, Ume\aa{}




\end{document}